\documentclass[graybox]{svmult}

\usepackage{type1cm}        
\usepackage{makeidx}         
\usepackage{graphicx}        
\usepackage{multicol}        
\usepackage[bottom]{footmisc}

\usepackage{newtxtext}       %
\usepackage[varvw]{newtxmath}       

\usepackage{subcaption}
\usepackage{placeins}

\makeindex             

\begin{document}

\title*{Simulating super-Chandrasekhar white dwarfs}
\author{Zenia Zuraiq\orcidID{0009-0000-6980-6334}, \\ Achal Kumar, \\ Alexander J. Hackett, \\ Surendra Bhattarai, \\ Christopher A. Tout\orcidID{0000-0002-1556-9449}, \\ Banibrata Mukhopadhyay\orcidID{0000-0002-3020-9513}}
\authorrunning{Zuraiq et al.}
\institute{Zenia Zuraiq \at Department of Physics, Indian Institute of Science, Bengaluru - 560012, India \email{zeniazuraiq@iisc.ac.in}
\and Achal Kumar\at Department of Physics, University of Florida,
Gainesville, Florida, USA \email{achal.kumar@ufl.edu}
\and Alexander J. Hackett \at Institute of Astronomy, University of Cambridge, Madingley Road, Cambridge CB3 0HA, UK and \\  CEICO, Institute of Physics of the Czech Academy of Sciences, Pod Vodárenskou věží 1999/1, 182 00 Praha 8-Libeň, Czechia \email{ajh291@cam.ac.uk, hackett@fzu.cz}
\and Surendra Bhattarai \at Department of Physical Sciences, Indian Institute of Science Education and Research, Kolkata, India \email{sb19ms136@iiserkol.ac.in}
\and Christopher A. Tout \at Institute of Astronomy, University of Cambridge, Madingley Road, Cambridge CB3 0HA, UK \email{cat@ast.cam.ac.uk}
\and Banibrata Mukhopadhyay \at Department of Physics, Indian Institute of Science, Bengaluru - 560012, India \email{bm@iisc.ac.in}}
%
%
\maketitle
  \textit{To be published in Astrophysics and Space Science Proceedings, titled "The Relativistic Universe: From Classical to Quantum, Proceedings of the International Symposium on Recent Developments in Relativistic Astrophysics", Gangtok, December 11-13, 2023: to felicitate Prof. Banibrata Mukhopadhyay on his 50th Birth Anniversary", Editors: S Ghosh \& A R Rao, Springer Nature}

\pagebreak


\abstract{Over the last few decades, there has been considerable interest in the violation of the {\it sacred} ``Chandrasekhar" mass limit of white dwarfs (WDs). Peculiar over-luminous type Ia supernovae (such as SNLS-03D3bb) lend observational support to the idea that these super-Chandrasekhar WDs exist. Our group, for more than a decade, has been actively working on the theoretical possibility of these objects through the presence of the star's magnetic field. The magnetic field greatly contributes to the existence of these massive WDs, both through classical and quantum effects. In this work, we explore super-Chandrasekhar WDs, formed via evolution from a main sequence star, as a result of the classical effects of the star's magnetic field. We obtain super-Chandrasekhar WDs and new mass limit(s), depending on the magnetic field geometry. We explore the full evolution and stability of these objects from the main sequence stage through the one-dimensional stellar evolution code STARS. In order to do so, we have appropriately modified the given codes by introducing magnetic effect and cooling. Our simulation confirms that massive WDs are possible in the presence of a magnetic field satisfying underlying stability. }

\section{Introduction}
\label{sec:1}
Type Ia supernovae (SNe Ia) have long been an integral part of the cosmic distance ladder, as standard candles used for measuring distances to far-away galaxies. It was the measurements using a population of SNe Ia that led to the Nobel prize winning discovery of the Universe's accelerated expansion. 

However, over the last few decades or so, there have been observations that indicate that these standard candles are not so standard after all. Specifically, there have been multiple observations of over-luminous SNe Ia, which have higher luminosities than the expected standard for these objects, along with a varied shape of the lightcurve and low ejecta velocity. These observations indicate that the progenitors of these objects are massive white dwarfs (WDs), that exceed the conventional Chandrasekhar limit, with some even as massive as $2.8M_\odot$ \cite{sne1a}. These are the proposed ``super-Chandrasekhar" WDs.

Mukhopadhyay and group have investigated the possible existence of these objects through various theoretical studies over the last decade or so (starting with the paper \cite{kundu}). In one of the series of avenues, they have found that the magnetic fields of WDs can lead to them having super-Chandrasekhar masses. 

Magnetic fields can lead to quantum effects on the WD structure through Landau quantization effects, as explored in previous work \cite{udas}. On the other hand, classical effects on the WDs through magnetic pressure and tension effects \cite{deb} can also lead to effects on the WD's mass. The consequence of these magnetic effects on the WDs is that instead of having a single Chandrasekhar mass limit, we have a series of mass limits, depending on the exact magnetic field included in the star. Nevertheless, they may have an ultimate (upper) mass limit, when we consider the star to be free from all instabilities.

In this work, we continue exploring magnetized WDs (BWDs), and more specifically, their formation and time-dependant evolution. We explore in detail the time-dependent simulation for the evolution of the magnetized main sequence (MS) stars to BWDs using the Cambridge Stellar evolution code: STARS \cite{eggleton}. In the original STARS, there were no magnetic fields. We have introduced fields in it by means of a profile of magnetic fields, which is a valid assumption as long as the star is approximately spherical. 

There are two ways in which this magnetic effect could affect the star's evolution. BWDs may be massive enough through their formation/evolution from the MS star itself. This may be possible when we consider the evolution of higher mass stars ($\approx$ 10$M_\odot$), which are sufficiently magnetized during the MS phase itself. An alternate formation scenario is possible when we consider binary systems. A WD with a low or dormant magnetic field may accrete matter (perhaps from a binary companion), leading to strengthening of its field and a higher mass limit for the WD. In the current work, we will be focussing on the second formation scenario, i.e., the ``accretion cases".


This paper is structured as follows. In Sec. \ref{sec:2}, we introduce the STARS code. We introduce the magnetic field profile and the modifications to the set of stellar structure equations to include magnetic fields in the simulation. In Sec. \ref{sec:3}, we highlight the major results, including the evolutionary track from the MS star to BWD and the properties of the BWDs obtained. We also investigate the possibility of how the evolution is affected by the decay of the magnetic field. In Sec. \ref{sec:4}, we lay out the future directions of this work. We end by summarizing our results and with our conclusions in Sec. \ref{sec:5}. 

\section{STARS Code: Formalism}
\label{sec:2}
The goal of this work is to explore the formation and evolution of BWDs and, by extension, super-Chandrasekhar WDs. For this, we have made use of the one-dimensional stellar evolution code, first developed at Cambridge, known as STARS. \cite{eggleton}. We evolve MS stars and study this evolution through various phases like the asymptotic giant branch (AGB) up to its formation as a BWD. 

Throughout our simulations, we have the following assumptions and considerations: 
\begin{enumerate}
    \item the star is spherical and one-dimensional,
    \item the star is nonrotating,
    \item we do not solve the hydrostatic balance and Maxwell's equations simultaneously, rather solve 
the hydrostatic balance equation with a modification to the pressure arising due to the model magnetic field profile.
\end{enumerate}

The stellar structure equations used in our simulation (in STARS), keeping the same symbols as used in the original STARS code, are:
\begin{equation}
    \frac{d\log P}{dm} = -\frac{Gm}{4\pi r^4P},
\end{equation}

\begin{equation}
    \frac{d\log T}{dm} = \frac{d\log P}{dm}\nabla,
\end{equation}

\begin{equation}
    \frac{d \log r}{dm} = \frac{1}{4\pi r^3m},
\end{equation}
and
\begin{equation}
    \frac{dL}{dm} = \epsilon - \epsilon_\nu - \frac{Du}{Dt} + \frac{P}{\rho^2}\frac{D\rho}{Dt}.
\end{equation}
Here, $\epsilon$ is the energy generation rate due to hydrogen burning, $\epsilon_\nu$ is the energy loss rate from neutrino emission, and $\nabla$ represents the logarithmic gradient of temperature against pressure, which varies for convective and radiative regions.

\subsection{Inclusion of magnetic field}
In our simulations, the magnetic field is calculated at every point in the star using the density-dependent profile \cite{bandyopadhyay} 

\begin{equation}
\label{prof}
B(\rho) = B_s + B_0\left[1 - \exp\left\{-\eta{\left(\frac{\rho}{\rho_0}\right)^\gamma}\right\}\right],
\end{equation}
where $B_s$ corresponds to the surface field of the star, $B_0$ and $\rho_0$ control the field at the center, and $\eta$ and $\gamma$ are
model parameters that control how the field decays from
the center to surface. In this work, we take $B_s = 10^7 \ G, B_0 = 10^{14} \ G \text{ and } \rho_0 = 10^9 \ g/cm^3$.

We incorporate magnetic fields by modifying the pressure to have contributions from both the matter ($P_m$) and the magnetic field ($P_B = B^2/8\pi$). 
We consider the star to have a toroidal field, leading to $P=P_m + P_B$. Note here that the magnetic density $\rho_B$ is throughout significantly less than the mass density for the fields under consideration. This form of magnetised pressure modification was followed in earlier work, where it was also shown that this form of the profile is consistent with Maxwell's equations under the assumption of approximate spherical symmetry. \cite{deb}. Further, toroidal fields lead to minimal deviation as shown in previous (time-independent) two-dimensional results from XNS code, e.g. \cite{surajit}.

\section{Results}
\label{sec:3}

\subsection{Formation of BWD models}

As mentioned previously, our goal is to simulate the complete formation and evolution of BWDs from the MS stars. For this, we have modified the STARS code to include magnetic field. We follow a similar implementation to form the WD as described in \cite{mukul}. The general scheme for the formation of a carbon-oxygen (CO) WD through the code is described below.

We use the STARS code to generate a zero-age main sequence (ZAMS) model. In the case of a CO WD, we would need to start with a MS model of mass $\approx 3-8 M_\odot.$
We evolve this star from the MS up to the AGB stage until its CO core has grown to a sufficiently large amount. 
At this point, all nuclear reactions in the star (chemical evolution) are halted, and a mass-loss mechanism in terms of the code's stellar wind parameter is enabled. This strips the outer layers of the star, leaving us with a mostly degenerate star. We thus obtain a CO WD. To obtain WDs of other compositions, one must start with appropriate ZAMS/MS models. An oxygen-neon WD would require a higher mass ZAMS model, such that conditions are favorable for carbon burning in the core. On the other hand, to form the lighter helium WDs, one must start with a sufficiently low mass ZAMS model, such that the core mass is not sufficient enough for the formation of a CO core in the first place. 

In this work, we assume the magnetic field effects are minimal in the MS phase. It is only on the star's collapse and field's growth (through either a flux freezing mechanism or perhaps some kind of dynamo effect) that the magnetic effects become important. 

Including a large magnetic field (either by $B_0$ or $B_s$) at this late stage of the star's evolution can lead to destabilizing effects. Keeping that in mind, the field is often introduced in stages so that the model has sufficient time to relax to the final BWD stage. The scale of this extra relaxation time is much smaller than the typical timescale of cooling or other subsequent processes (such as accretion/mass addition) introduced to the WD.

\begin{figure}[!htpb]
	\centering
	\includegraphics[scale=0.65]{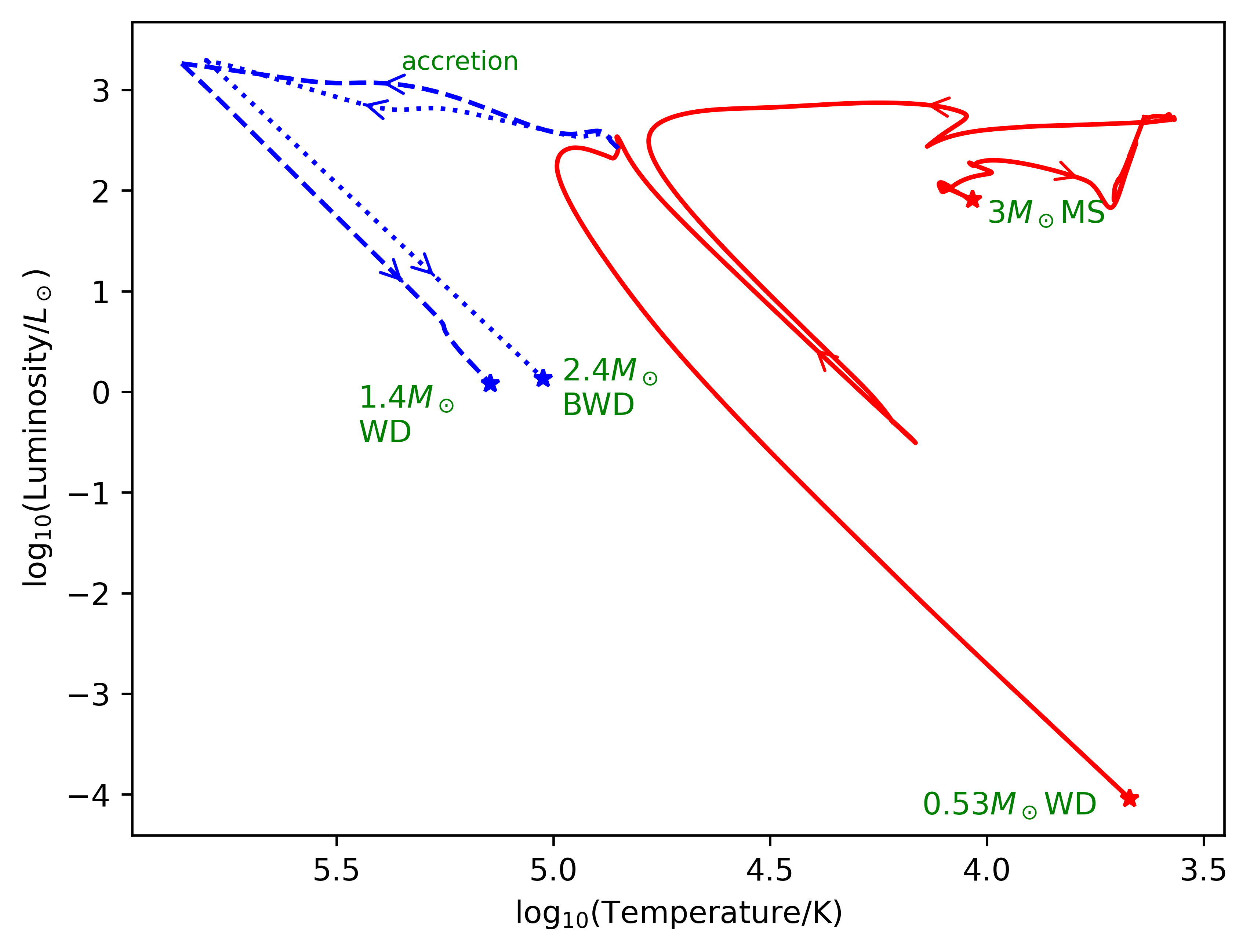}
	\caption{Evolution of MS star to BWD as shown by changing position on the HR diagram. The subsequent cooling track of BWD is not shown here.}
    \label{hr}
\end{figure}

Here, we focus on a particular evolution scenario: the formation of a $0.4 - 0.5 M_\odot$ BWD formed from a $3 M_\odot$ MS star. We can visualize the various stages of the evolution process towards the formation of the BWD on the Hertzsprung-Russell (HR) diagram, as shown in Fig. \ref{hr}. The solid red line indicates the formation of $0.53M_\odot$ BWD. Following this, we consider accretion onto the WD (described in detail in next subsection). Accretion onto a WD having magnetic field is shown with the dotted line, while the dashed line shows accretion onto a non-magnetic WD. Depending on whether the field is there or not, the evolution ends up with different limiting masses for the BWDs.

\subsection{Super-Chandrasekhar WDs: Series of mass limits for a BWD}
As described in the previous subsection, through the generation and subsequent evolution of an appropriate mass ZAMS model, one can evolve a CO WD using STARS. Further, on introducing the pressure modification by the magnetic field, one obtains a BWD model.

The magnetic field of a BWD can lead to interesting effects. As mentioned previously, the magnetic field leads to a modification of the conventional Chandrasekhar limit through both classical and quantum effects. In our modification, we have focussed on the modification to the hydrostatic equilibrium (pressure balance) through the inclusion of a toroidal magnetic field in the star.

To examine this change of mass and mass limit through our simulation, we need to examine the maximum limit of how much mass our BWD can support against gravity. We introduce accretion to our star, through the code's (spherical) mass addition control. By adding mass to our BWD model, we obtain higher and higher mass BWDs. Higher mass BWDs have denser cores, and thus higher fields at the center as well. Thus, along with the growth of mass in subsequent models, one can also see the growth of the central field of the BWD.

\begin{figure}[!htpb]
	\centering
	\includegraphics[scale=0.55]{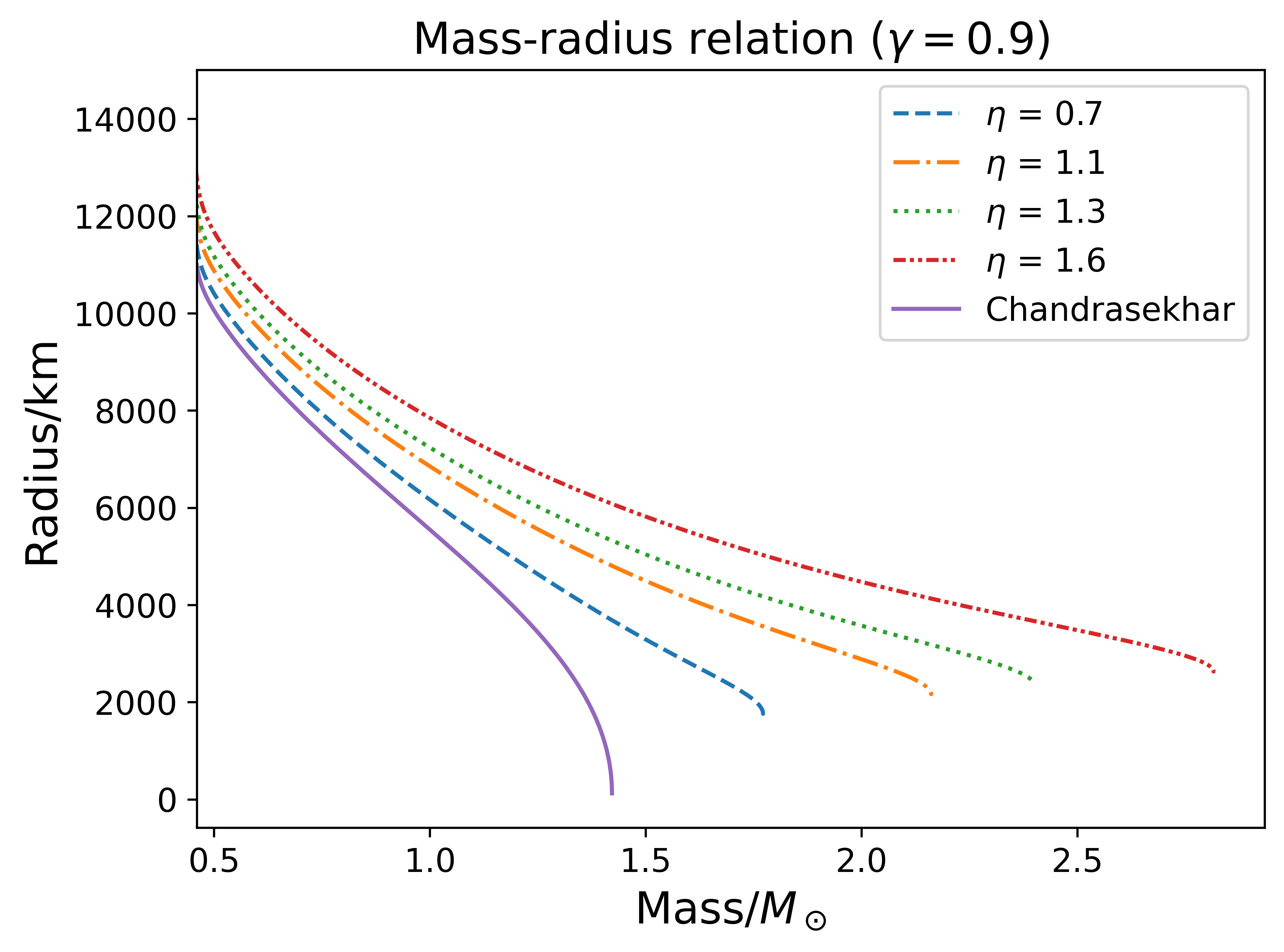}
	\caption{M-R curves showing different mass limits from different magnetic field profiles (different values of $\eta$). The mass accretion rate is $10^{-9}M_\odot$/Yr. }
    \label{mr1}
\end{figure}

One can examine the change in mass limit through the various mass-radius (M-R) curves for the WDs obtained by mass addition. By comparing the M-R curves obtained for magnetized and non-magnetized cases, we see that there is a clear enhancement of saturation mass in the former, as shown in Fig. \ref{mr1}.  The mass limit is now much higher than the Chandrasekhar limit - in one case, it is as high as $2.8M_\odot$. These are the theorized super-Chandrasekhar WDs, finally realized in a time-dependent simulation,
via evolution from the MS star! 

Further, as shown in the figure, different magnetic field profiles (represented here through the change of $\eta$) lead to different saturated masses/mass limits. It is not just the strength of the magnetic field, but the exact profile/geometry that it follows within the star, that finally determines the exact mass limit obtained.

\begin{figure}[!htpb]
	\centering
	\includegraphics[scale=0.55]{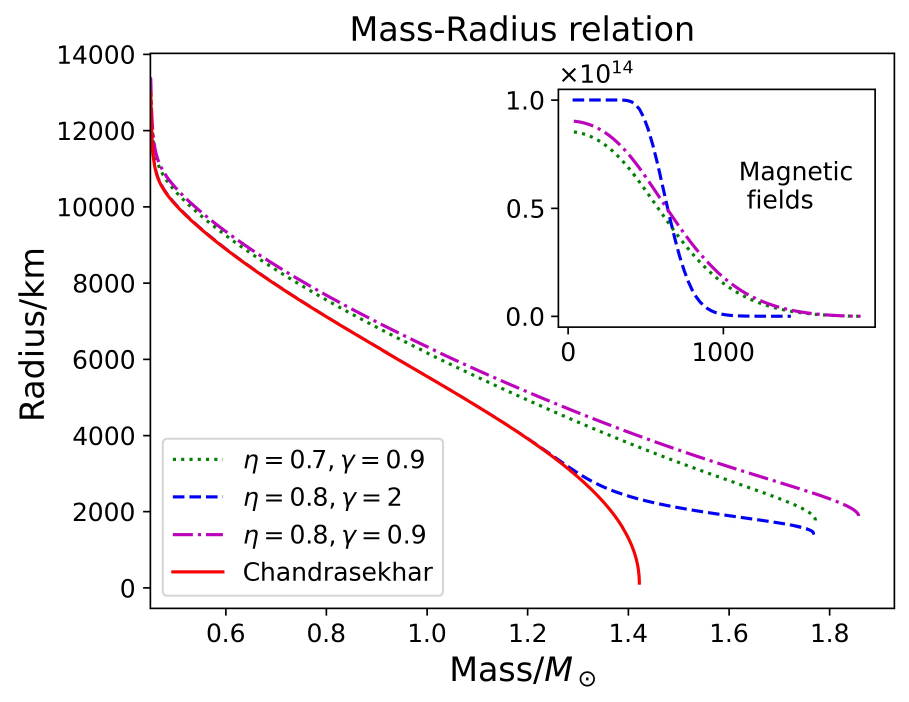}
	\caption{M-R curves showing different radii in the low mass (sub-Chandrasekhar) regime based on different profiles.}
    \label{mr2}
\end{figure}

Fig. \ref{mr2} shows another interesting consequence of the change of field profile. While both the field profiles considered here give the same saturated mass/mass limit, due to the difference in their gradients, their M-R curves are different. The blue dashed curve ($\eta = 0.8, \gamma = 2$) and the green dotted curve ($\eta = 0.7, \gamma = 0.9$) have the same limiting mass, but the green dotted case gives higher radii at lower masses. This shows how the effect of magnetic field could be seen in sub-Chandrasekhar mass WDs in specific cases, and this could ultimately have important implications for observations.


Thus, in the presence of magnetic field, one obtains a series of mass limits for a WD, depending on the exact field strength and geometry introduced. This was earlier shown analytically by \cite{deb} and we are now able to reproduce the result through time-dependent simulations as well. Further, the magnetic field can affect sub-Chandrasekhar WDs, and this could be an important observational discriminant.

We are able to establish the accretion formation scenario for super-Chandrasekhar WDs, where an initially lowly magnetized WD accretes matter, leading to the field of the star being strengthened, e.g. by flux conservation. Eventually, accretion can lead to a maximal mass beyond the Chandrasekhar limit. As we show here, the exact maximum mass is very much dependent on the details of the magnetic field present within the star.




\subsection{Magnetic field decay}

Given that we are examining how the magnetic field can lead to a WD supporting super-Chandrasekhar masses, it is interesting to examine what happens when this field changes. Particularly, we know that the magnetic field of stars decays over time through a combination of different processes. A natural question to ask then is: \textit{What happens to the extra mass, which was held due to the stronger field, when the field decays?} Two possible scenarios could arise here - the star may shed the extra mass through some mechanism and end up as a Chandrasekhar WD; or the star may collapse, unable to support the extra mass against gravity in absence of the field. The latter scenario points towards a SNe type evolution.

There are three main processes that contribute to field decay in compact stars: ambipolar diffusion, Ohmic decay, and Hall effect. \cite{goldreich}. In WDs, generally ambipolar diffusion does not feature, and it is the other two processes which are dominant. Field decay by a combination of Ohmic and Hall effects was explored analytically in previous work by our group \cite{mukul}.

Following previous work \cite{heyl}, we assume the magnetic field decay to take the following form:
\begin{equation}
    \frac{dB}{dt} = -B\left(\frac{1}{t_{Ohm}}+\frac{1}{t_{Hall}}\right),
\end{equation}
where $B$ is the magnitude of magnetic field, as calculated by eqn. \ref{prof}. We use the expressions for the characteristic time-scales of these equations from previous analytical work \cite{mukul}, given by
\begin{equation}
    t_{Ohm} = (7 \times 10^{10}) \rho_{c,6}^{1/3}R_4^{1/2}\frac{\rho_{av}}{\rho} \text{years,}
\end{equation}

\begin{equation}
    t_{Hall} = (5 \times 10^{10}) l_8^2 B_{0,14}^{-1} T_{c,7}^2\rho_{c,10} \text{years,}
\end{equation}
where $\rho_{c,n} = \rho_c/10^n \ g/{cm}^3$ , $\rho_c$ is the central density of the WD, $R_{4} = R/10^4 \ km$, $T_{c,7} = T_c/10^7 \ K$, $B_{0,14} = B_0/10^{14} \ G$ and $l = l_8 \times 10^8 \ cm$ is characteristic length scale of the flux loops through the outer core of the WD.

Given the constraints of a one-dimensional hydrostatic stellar evolution code, we cannot hope to capture the entirety of either of these extremely dynamic processes. However, as a preliminary analysis, one can introduce field decay as one of the set of equations to be solved. We can then examine what scenario the code points us to. 

\begin{figure}[!htb]
     \centering
     \begin{subfigure}{0.475\textwidth}
         \centering
         \includegraphics[width=0.95\textwidth]{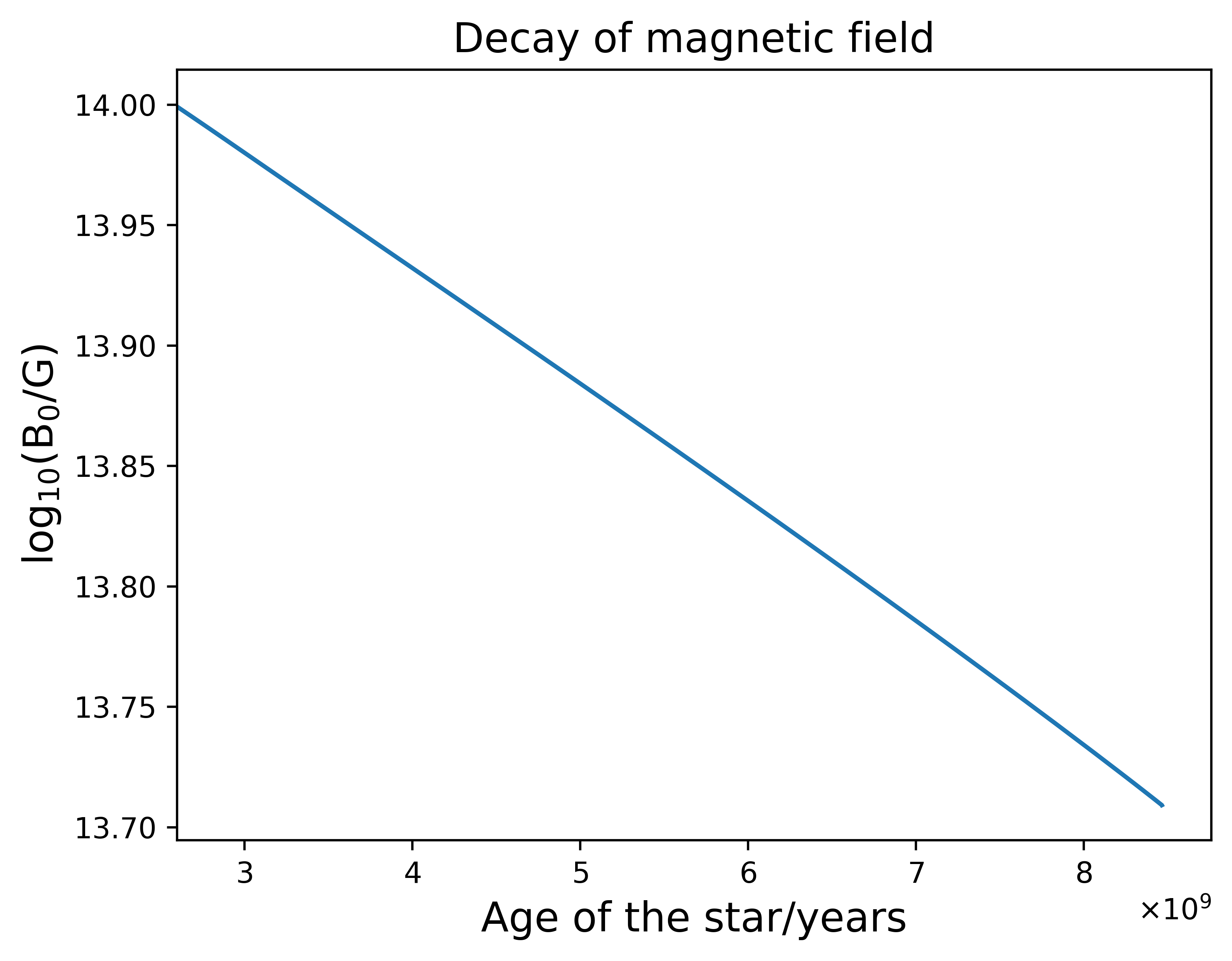}
     \end{subfigure}
     \hfill
     \begin{subfigure}{0.475\textwidth}
         \centering
         \includegraphics[width=0.95\textwidth]{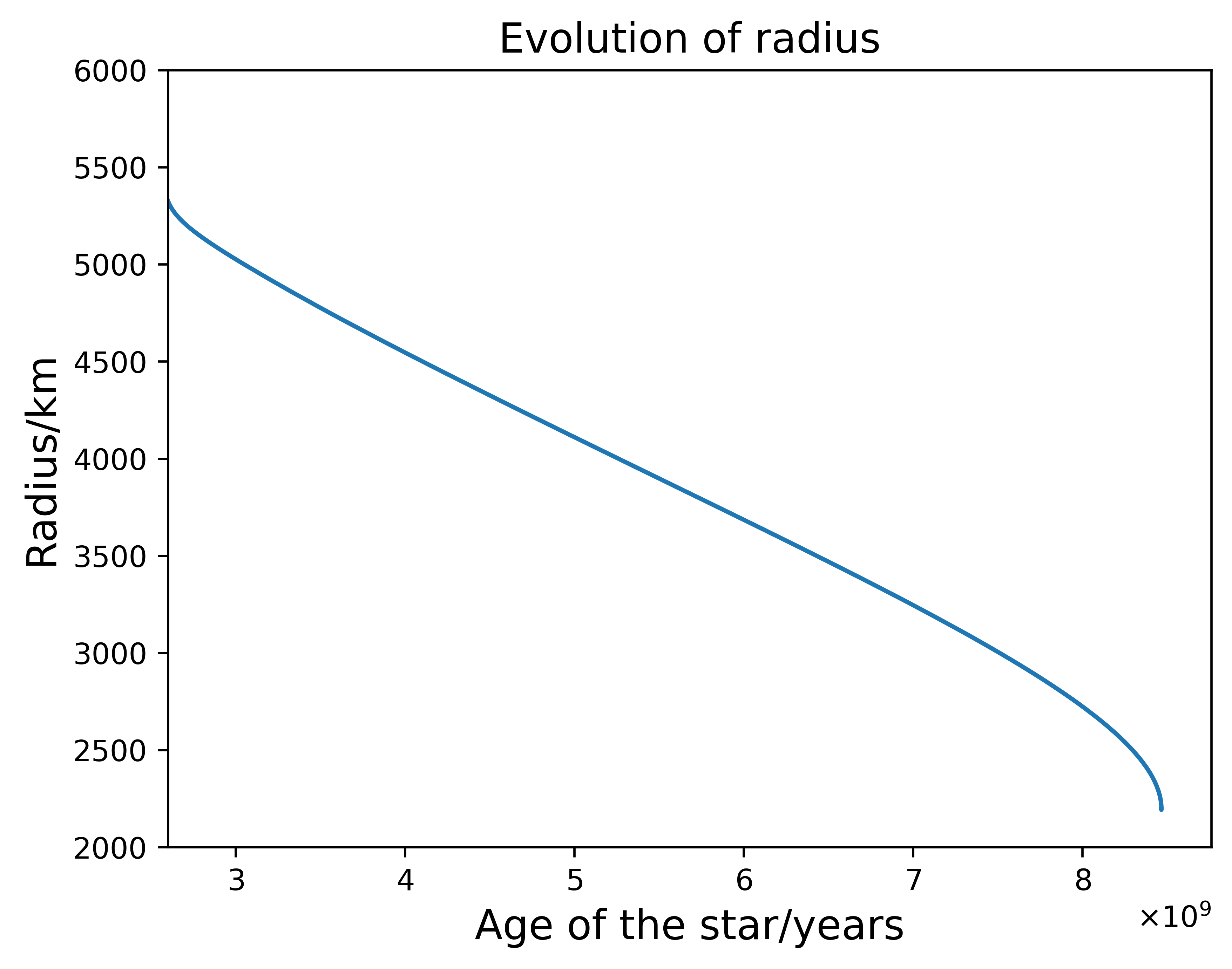}
     \end{subfigure}
     \vskip\baselineskip
          \begin{subfigure}{0.475\textwidth}
         \centering
         \includegraphics[width=0.95\textwidth]{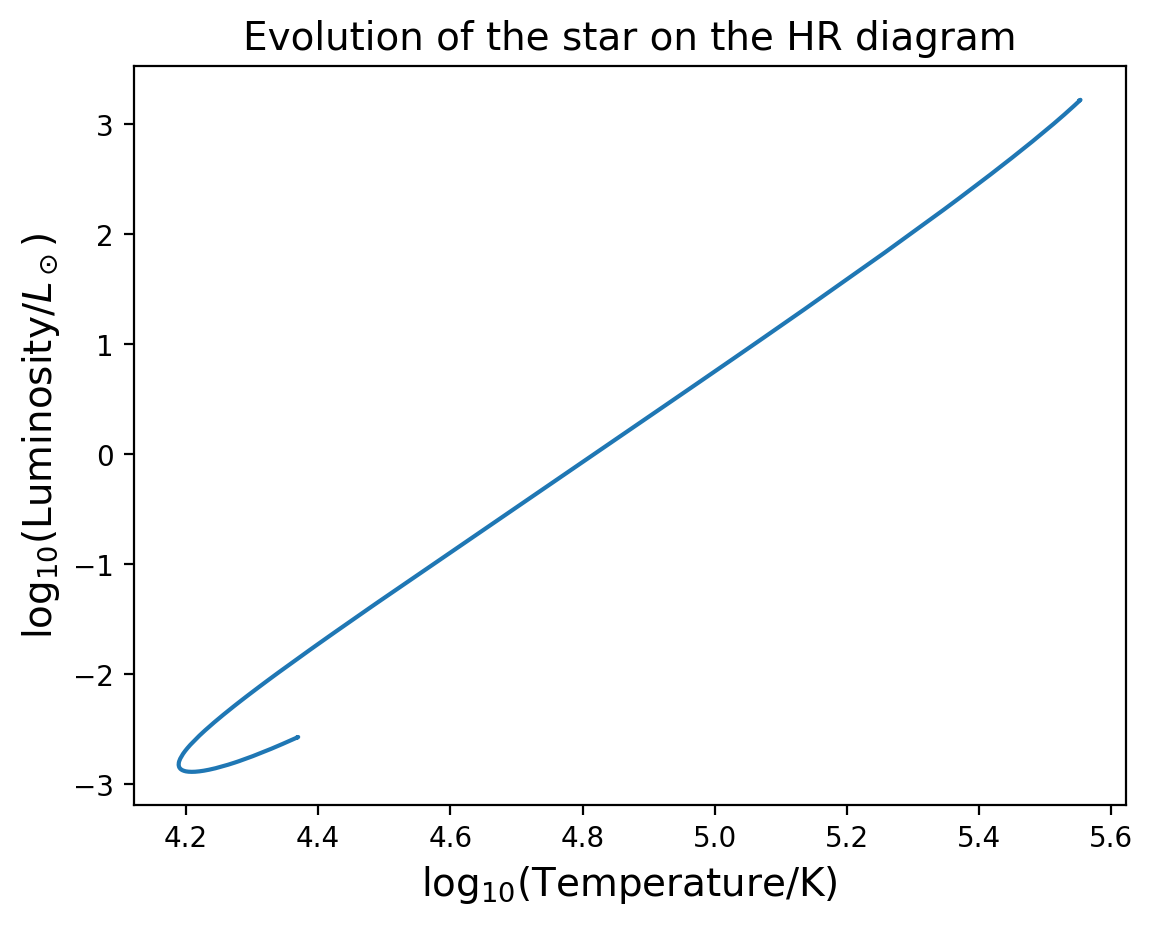}
     \end{subfigure}
     \hfill
     \begin{subfigure}{0.475\textwidth}
         \centering
         \includegraphics[width=0.95\textwidth]{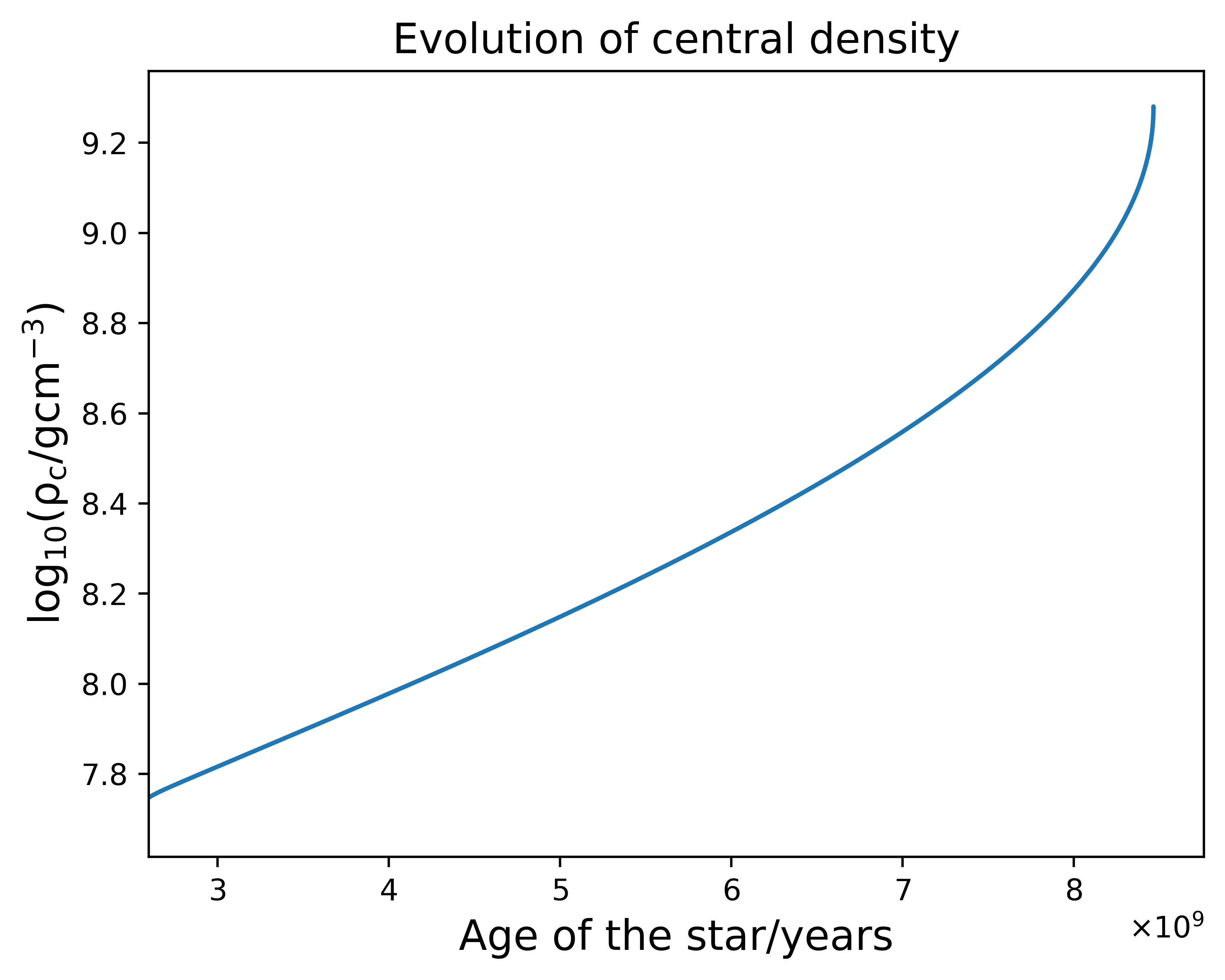}
     \end{subfigure}
     \centering
     \captionsetup{justification=raggedright}
     \caption{Change in various stellar properties as the magnetic field decays.}
        \label{decay}
\end{figure}

We find some interesting results before the code crashes (due to triggering of instability). As seen from the top panels of Fig. \ref{decay}, the decay of the magnetic field is accompanied by a contraction of the star, seen by its decrease of radius. The WD becomes nearly half its initial radius before the instability triggers. Further, as shown in the lower panels of Fig. \ref{decay}, this is also accompanied by a simultaneous increase in the density and luminosity (in the last phase) of the star. Both the density and luminosity increase sharply just before the run stops. These are all indications that the WD is moving towards a denser phase of evolution, i.e., these all indicate the WD collapsing to form a SNe Ia. This result is not altered qualitatively on varying the decay rate of the magnetic field.

Thus, our preliminary analysis seems to argue the fact that SNe Ia do not necessarily originate only from the thermonuclear explosions of Chandrasekhar mass WDs, and may be formed from the subsequent evolution of super-Chandrasekhar BWDs. 
\FloatBarrier
\section{Future (and on-going) Work}
\label{sec:4}

We have described here a set of preliminary results showing the time-dependent simulation of the formation and evolution of BWDs, and their further evolution to super-Chandrasekhar BWDs. These initial results are extremely promising, as they already point us towards the stability and realisability of these objects.

We have looked at parallel results from another stellar evolution code, MESA \cite{MESA}. We find consistency with formation mechanisms in the non-magnetic cases. Work is underway to incorporate magnetic fields in a similar fashion in MESA. MESA also allows for more realistic accretion, with controls present to specify the composition of accreted matter.

We are currently working towards making results more robust in a number of ways. For instance, we are currently including the chemical evolution through nuclear reactions throughout the WD phase as well. Although there is minimal burning in the WD once it is fully degenerate, there are cases where sufficient burning of the envelope can lead to some interesting results. We are also examining the effects of having a more realistic mixed-field geometry in our star, as it is long known that purely toroidal and purely poloidal fields are magnetohydrodynamically unstable \cite{braithwaite}. These changes to the implementation do not majorly change the qualitative results. Lowly magnetized/dormant field based BWDs can accrete to form super-Chandrasekhar BWDs, even with all these caveats considered.



Finally, we also examine the alternate formation scenario for super-Chandrasekhar WDs mentioned earlier. BWDs may be massive enough through their formation/evolution from a sufficiently massive and magnetized MS star itself.

\section{Summary and Conclusions}
\label{sec:5}
The Chandrasekhar limit is one of the fundamental results of theoretical, stellar and high energy astrophysics. It brings together astrophysics (cosmology), quantum statistics and gravity in a single, beautiful formula. However, observations of over-luminous SNe Ia tell us that it is not the complete picture. Particularly, effects due to finite temperature, rotation, and magnetic fields were not considered in Chandrasekhar's treatment. All these effects, separately and together lead to clear theoretical deviations from the ideal Chandrasekhar M-R relation and mass limit. 

Through the current work, we capture the finite temperature effects in the WDs by solving for full stellar structure. Further, through modification of the code, we introduce magnetic effects as well. Through these effects, we are able to reproduce super-Chandrasekhar WDs through time-dependent simulations. We obtain the full evolutionary track from the MS star up to the formation of BWD. By accreting onto both magnetized and non-magnetized WDs, one clearly obtains different M-R relations, and thus different mass limits. Thus, we have now established a stable formation scenario for a super-Chandrasekhar WD, consisting of the inital formation of lowly magnetized BWD, whose field is further enhanced through accretion, leading to it supporting masses beyond the Chandrasekhar limit. This is consistent with the observational fact that a higher percentage (20-25\%) of WDs in binary systems are magnetic \cite{ferrario}. 
On doing a preliminary analysis of magnetic field decay in the star, we see that these super-Chandrasekhar WDs tend towards SNe type evolution, further cementing their link to observations of overluminous SNe Ia.

Our work is definitive proof that the super-Chandrasekhar BWDs, proposed a decade ago by Mukhopadhyay and group, are stable and realizable. Through our simulation, we are able to establish a time-dependent evolutionary track showing the formation of these objects, with some tracks ending in WDs with masses as high as $2.8M_\odot$. This is probably the first work of its kind, and the first successful STARS simulation which shows such highly super-Chandrasekhar WDs. Our results have far–reaching implications which may even caution against SNe Ia as standard candles.

\begin{acknowledgement}
Z. Z. thanks the organizers of the International Symposium on Recent Developments in Relativistic Astrophysics (ISRA) for the opportunity to present this work. Z. Z. also acknowledges the Prime Minister’s Research Fellows (PMRF) scheme, with Ref. No. TF/PMRF-22-7307, for providing fellowship.
\end{acknowledgement}
\ethics{Competing Interests}{
The authors have no conflicts of interest to declare that are relevant to the content of this chapter.}

\end{document}